\documentclass[aps,prl,twocolumn,nobibnotes]{revtex4-1}
\usepackage{graphics,graphicx,amsfonts,amsmath,amsbsy,amssymb,color}
\usepackage{bm}
\usepackage[a4paper,vmargin={20mm,20mm},hmargin={20mm,10mm}]{geometry}
\usepackage{subfigure}
\usepackage{paralist}
\usepackage{bbold}

\def \beq {\begin{eqnarray}}
\def \eeq {\end{eqnarray}}

\def \Lowdin {{L\"{o}wdin }}

\def \nadd {{n_{\textrm{a}}}}
\newcommand{\bra}{\langle}
\newcommand{\ket}{\rangle}

\begin{document}
\title{Krylov-projected quantum Monte Carlo}
\author{N.~S.~Blunt}
\email{nsb37@cam.ac.uk}
\author{Ali~Alavi}
\author{George~H.~Booth}
\email{george.booth@kcl.ac.uk}
\affiliation{University Chemical Laboratory, Lensfield Road, Cambridge, CB2 1EW, United Kingdom}
\affiliation{Max Planck Institute for Solid State Research, Heisenbergstra{\ss}e 1, 70569 Stuttgart, Germany}
\affiliation{Department of Physics, King's College London, The Strand, London, WC2R 2LS, U.K.}

\begin{abstract}
We present an approach to the calculation of arbitrary spectral, thermal and excited state properties within the full configuration interaction quantum Monte Carlo framework. 
This is achieved via an unbiased projection of the Hamiltonian eigenvalue problem into a space of stochastically sampled Krylov vectors, thus enabling the calculation of 
real-frequency spectral and thermal properties and avoiding explicit analytic continuation. We use this approach to calculate 
temperature-dependent properties and one- and two-body spectral functions for various Hubbard models, as well as isolated excited states in {\em ab initio} systems.
\end{abstract}
\date{\today}
\maketitle

Quantum Monte Carlo (QMC) in its various guises, is undoubtedly one of the most important approaches for accurate
elucidation of properties for correlated systems\cite{Foulkes2001,Booth2013,Sharma2014,Zhang2012,SorellaSpinLiquid}.
However, these successes have focused primarily on the 
ground state energy and observables which commute with the Hamiltonian. Critical importance for a deeper understanding of 
correlated systems comes from {\it dynamic} correlation functions and spectral quantities. 
These mirror how we perceive our environment, namely by perturbing a system and measuring its response -- the basis of nearly all spectroscopic and experimental approaches. This gives us direct insight into optical, magnetic and other beyond-ground-state properties, and allow for direct comparison to experimental results.

Direct access to dynamic properties is a persistent difficulty for QMC approaches in general. While in the absence of a sign
problem, unbiased imaginary-time spectra can be obtained\cite{White1989_2,Motta2014,SSE}, the analytic continuation to
physical, real-frequency functions is notoriously ill-conditioned and can lead to artefacts and smoothing of features.\cite{Jarrell1996}
For more general Fermionic systems, 
higher temperatures must be simulated to alleviate the sign problem\cite{Gull07}, while nodal constraints bias 
towards a particular solution and are difficult to extend to spectra\cite{Motta2014,Foulkes2001,Williamson1998}.
Alternatively, projections into effective Hamiltonians
have been able to obtain a few low-energy states, but again these are isolated states rather than practical approaches for thermal or 
spectral quantities\cite{Ceperley1988,Tenno2013}, while a modification of the propagator can lead to debilitating timestep issues\cite{Chan_H2}.

Here, we present a new QMC approach for computing dynamic correlation functions, temperature-dependent quantities and isolated excited states for correlated quantum systems, even in the presence of a sign-problem. 
These correlation functions
are unbiased in the limit of large averaging, and exact in the limit of large walker number.
This is achieved by extending the recently developed 
Full Configuration Interaction Quantum Monte Carlo (FCIQMC) method\cite{Booth2009,Booth2013,Spencer2012}, by combining it with ideas from the 
dynamical and finite-temperature Lanczos (FTLM) methods.\cite{FTLM,Prelovsek2013,MPSLanc} The key advantage of the approach is that it avoids any explicit storage over the full Hilbert space, instead only storing occupied states in the discretized wavefunction at each snapshot. This allows for sparsity in the wavefunction to be exploited to minimize memory bottlenecks, which are a primary limitation in conventional approaches which require explicit storage over the space\cite{FTLM,Prelovsek2013,Hams2000,Shimizu2013}. 
The result is a QMC method which although weakly exponentially scaling, in common with the ground state FCIQMC approach, can allow for systems to be treated well outside that possible by conventional means, and retains many of the important features of the parent method\cite{Booth2009,shepherd2014}. These include a cancellation algorithm to ameliorate the sign problem, an absence of time-step error and large-scale parallelism.

An arbitrary dynamic correlation function is defined as
\begin{equation}
    G(\omega)=\bra \Psi_0 | {\hat A}^{\dagger} \frac{1}{\omega - ({\hat H} - E_0) + i\eta} {\hat V} | \Psi_0 \ket,
\label{eq:correlation_function}
\end{equation}
where ${\hat H}$ represents the Hamiltonian of the system, $\{|\Psi_0\ket; E_0\}$ is the ground state wavefunction and energy, $\eta$ is a small broadening
parameter,
and ${\hat V}$ and ${\hat A}$ are arbitrary operators which define the perturbation and observed quantity in 
the correlation function. In the case of these operators being single annihilation and creation operators one 
obtains the single-particle Green function, 
defining the system bandstructure and density of states. 

The aim of our method 
is to stochastically obtain a projection of the Hamiltonian from the complete Hilbert space to an effective, reduced dimensionality space, such that 
it spans the degrees of freedom required to accurately describe the desired spectral or thermal quantity. In this work, we use a set of stochastically
sampled wavefunctions from a FCIQMC calculation to define this transformation. If the initial state of the calculation is a stochastic 
representation of the wavefunction ${\hat V}|\Psi_0\ket$, then propagation from this state\cite{fn1} to the ground state will in principle span all states required
to represent the expression in 
Eq.~\ref{eq:correlation_function}, 
equivalent to the space of ground state and all imaginary-time response vectors. 
Once the Hamiltonian is projected into this space, 
it can be exactly diagonalized, and the
desired correlation function of Eq.~\ref{eq:correlation_function} directly constructed in this eigenbasis -- the Lehmann 
representation. For thermal quantities the approach is analogous, with the initial 
vector taken from the infinite-temperature distribution.
Similar themes have been explored within continuum QMC, but applied to accelerate convergence for ground state properties\cite{Caffarel1991}.

\emph{Method:-} An FCIQMC iteration consists of stochastically applying a projection operator, $\boldsymbol{P}$, to a walker distribution,
denoted at iteration $i$ by $\boldsymbol{q}_i$, such that exact projection is achieved on average, whose distributions we denote as $\boldsymbol{\psi}_i$\cite{fn1}.
The aim is to stochastically sample the Krylov 
subspace $ \left\{ \boldsymbol{\psi}_0, \boldsymbol{P} \boldsymbol{\psi}_0, \dots , \boldsymbol{P}^{n-1} \boldsymbol{\psi}_0 \right\}$. In projector QMC approaches one samples from the large $n$ limit of this subspace, which converges to the ground state. However, to obtain
finite-temperature and dynamic quantities, the aim is now to stochastically project the Hamiltonian into the whole sampled Krylov subspace, which represents
an efficient span of all states of interest, provided that $\boldsymbol{q}_0$ is chosen appropriately.

By averaging the FCIQMC walker amplitudes, 
the results of an exact propagation are rigorously approached for expectation values which depend linearly on the wavefunction\cite{Booth2009,Spencer2012}. 
In this work, quadratic quantities 
are required, but now $E[\boldsymbol{q}_{i}^{\dagger} \boldsymbol{q}_{j}] \neq \boldsymbol{\psi}_i^{\dagger} \boldsymbol{\psi}_j$, due to correlations between walker amplitudes, 
where $E[\boldsymbol{q}_i]$ denotes the expectation value. To compute these, two independent replica sets of walkers are propagated simultaneously (indexed via superscripts), such that the amplitudes are uncorrelated between them\cite{Zhang1993,Blunt2014}, allowing for unbiased estimates of $\boldsymbol{\psi}_i^{\dagger} \boldsymbol{\psi}_j$ as $E[\boldsymbol{q}_{i}^{1 \dagger} \boldsymbol{q}_{j}^2]$ or $E[\boldsymbol{q}_{i}^{2 \dagger} \boldsymbol{q}_{j}^1]$.
This approach for static correlation functions has been found to scale without difficulty within FCIQMC\cite{Overy2014}.

At selected iterations in an FCIQMC calculation the walker distribution is stored\cite{fn2}, and the 
overlap ($\boldsymbol{S}$) and Hamiltonian ($\boldsymbol{T}$) matrices between these subspace vectors calculated as
\begin{align}
S_{ij} &= (\boldsymbol{q}_{i}^{1 \dagger} \boldsymbol{q}_{j}^2 + \boldsymbol{q}_{i}^{2 \dagger} \boldsymbol{q}_{j}^1)/2, \\
T_{ij} &= (\boldsymbol{q}_{i}^{1 \dagger} \boldsymbol{H} \boldsymbol{q}_{j}^2 + \boldsymbol{q}_{i}^{2 \dagger} \boldsymbol{H} \boldsymbol{q}_{j}^1)/2.
\end{align}
Whilst the overlap matrix estimate is trivial, calculating the $\boldsymbol{T}$ matrix exactly is expensive, and so instead it is stochastically sampled in the same manner as spawning steps in FCIQMC\cite{Overy2014}. 
Thus, a simulation provides an estimate of the overlap matrix and the projected Hamiltonian in the basis of Krylov vectors chosen, and so we denote the method {\em Krylov Projected} (KP)-FCIQMC. Averaging these quantities over independent simulations can reduce errors in an unbiased manner, resulting in a generalised eigenvalue problem for the projected Hamiltonian. This can be solved by standard techniques (see supplementary material). 
Many of the eigenvalues of $\boldsymbol{S}$ will be very small (or even negative within stochastic errors), 
since the sampled space becomes increasingly linearly dependent with continued propagation. We therefore discard these vectors of $\boldsymbol{S}$ without substantial loss of information. We refer to the eigenvectors which are kept as \Lowdin vectors.
We note that although the estimates of $\boldsymbol{T}$ and $\boldsymbol{S}$ are unbiased, the final eigenvalues will not be because eigenvalues 
are non-linear functions of these matrices.
However, this bias can be systematically reduced with further averaging of $\boldsymbol{T}$ and $\boldsymbol{S}$.

For exact propagation with $\hat{P}=\hat{H}$, our approach will yield results identical to the Lanczos method. 
However, because the method exploits sparsity via a stochastic representation of the wavefunctions, large calculations can often use significantly less memory than an equivalent Lanczos calculation, as has been the case for ground-state FCIQMC.
Although our approach is in theory systematically improvable to exactness for the entire frequency range, in practice this becomes increasingly difficult for higher energy excitations.
This is because high-energy excitations have a small component in the Krylov vectors, which decreases exponentially with imaginary time. This renders them particularly difficult
to sample and susceptible to stochastic error in the sampled matrices.
Despite this limitation, the approach can nevertheless be expected to obtain near-exact spectra
for low-energy excitations in systems out of reach of traditional dynamical Lanczos approaches.

\emph{Finite-temperature:-} We assess the method with the half-filled Hubbard model (defined in supplementary information)\cite{Spencer2012,shepherd2014}. 
Within the FTLM, thermal expectation values are computed via
\begin{equation}
\textrm{Tr}(e^{-\beta \hat{H}} \hat{A}) = \sum_{n=1}^N \sum_{i=0}^{M-1} e^{- \beta E_i^n} \bra n | \psi_i^n \ket \bra \psi_i^n | \hat{A} | n \ket + \mathcal{O}(\beta^{M}),
\label{eq:thermal_approx}
\end{equation}
where 
$|n\ket$ labels a state in the $N$-dimensional Hilbert space, and $i$ labels the $M$ states of an eigensystem $\{|\psi_i^n \ket\ ; E_i^n\}$
resulting from a Lanczos subspace with initial state $|n\ket$. Thus, by performing $N$ Lanczos calculations consisting of $M-1$ applications of ${\hat H}$ each, one can obtain thermal quantities which are correct to order $\beta^{M-1}$. $N$ can be very large for systems of interest and so in practice one starts from a much smaller number of states, $R \ll N$, chosen as a random linear combination of all basis states, $|r\ket = \sum_n \eta_{rn}|n\ket$. This turns out to converge quickly with $R$, 
particularly at high temperatures\cite{FTLM,Prelovsek2013,Iitaka2004,Schnack2010,Hanebaum2014}. 
In our stochastic approach the initial random vectors are created by distributing a given number of walkers randomly throughout the Hilbert space with coefficients $\pm1$. 
These initial states represent stochastic snapshots of the high-temperature limit which is exactly reproduced in the limit of large $R$.

\begin{figure}
\includegraphics{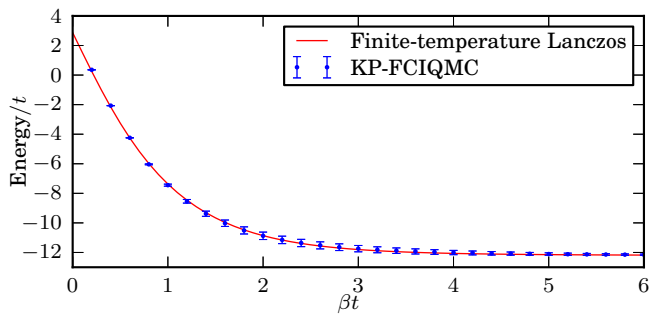}
\caption{$E(\beta)$ for the 12-site 1D Hubbard model at $U/t=1$ sampled with $\sim 2\times10^3$ walkers, with comparison FTLM. Error bars show standard deviation (note not standard error) over $10$ independent calculations to demonstrate the spread of results. High and low temperature results are almost exact, whilst at intermediate temperatures, the variance in the stochastic sampling as well as systematic errors (such as from the non-linear diagonalization step, and finite $R$) increases the variation between runs. Simulation parameters were $\tau=0.01$, $\nadd=2.0$, and a deterministic space 
of double excitations\cite{Petruzielo2012,EAs_iFCIQMC}.}
\label{fig:finite_t_12}
\end{figure}

As an initial test, Figure~\ref{fig:finite_t_12} presents the temperature-dependent energy, $E(\beta)$, in the one-dimensional 12-site Hubbard model at $U/t=1$. Including all symmetry sectors the Hilbert space dimension is $\approx 3 \times 10^6$, with the largest symmetry
sector containing $\approx 7 \times 10^4$ determinants. However, the system is significantly undersampled with only $2\times10^3$ walkers used throughout, with the projected Hamiltonian and overlap matrices averaged over $10$ calculations for each initial vector, $|r\ket$. All symmetry sectors were obtained in one calculation, rather than symmetry-blocking Eq.~\ref{eq:thermal_approx}, resulting in a choice of $R=1250$, while the number of Krylov vectors used was $M=20$, with $8$ \Lowdin vectors kept to form the final space. 
The results were found not to change significantly by including more \Lowdin vectors. 

At high temperatures results are calculated with great accuracy. This is easily understood because the quantity calculated at $\beta=0$, $\sum_{r=1}^R \sum_{i=0}^M \bra r | \hat{H} | \psi_i^r \ket \bra \psi_i^r | r \ket$, is rigorously equal to $\sum_{r=1}^R \bra r | \hat{H} | r \ket$, and therefore the quality is mainly dependent on the sampling of the initial vectors (and not on the error of individual eigenvalue estimates). 
At low temperatures the results are dominated by the ground state, which has a large component in the sampled Krylov vectors and so is accurately calculated by KP-FCIQMC. 
However, at intermediate temperatures the errors are larger. The most significant source of error is in replacing an exact trace over $\left\{|n\ket\right\}$ by an approximate one over $\left\{|r\ket\right\}$ in Eq.~\ref{eq:thermal_approx}.

\begin{figure}
\includegraphics{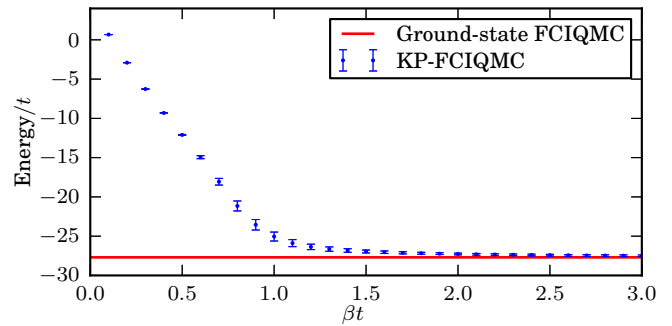}
\caption{$E(\beta)$ for the 18-site 2D Hubbard model at $U/t=1$, with ground-state FCIQMC energy for comparison. 10 independent simulations were used to create the standard deviations shown as error bars. Simulation parameters were $\tau=0.01$, $\nadd=2.0$, and a deterministic space of double excitations.} 
\label{fig:finite_t_18}
\end{figure}

In Figure~\ref{fig:finite_t_18}, $E(\beta)$ for the two-dimensional 18-site Hubbard model at $U/t=1$ is presented. Including all symmetry sectors the Hilbert space dimension is $\approx 9\times10^9$, with the largest symmetry sector containing $\approx 1\times10^8$ determinants. Again, the space was undersampled, with $5 \times10^6$ walkers used throughout, with $R=250$ and $M=20$, of which $12$ \Lowdin vectors are kept. 
Since FTLM was unfeasible, also plotted is a highly-accurate ground-state FCIQMC energy for comparison. A complete calculation took around $\sim3000$ core hours. 
We find again that the high-temperature results have only a small variation between repeated calculations and we have a high degree of confidence here. At lower temperatures the confidence in the results is reduced, with possible systematic errors including initiator error, bias in the eigenvalue estimates and an insufficient choice of $R$.

\emph{Dynamical correlation functions:-} To demonstrate the ability of KP-FCIQMC to calculate dynamical quantities, we first consider the following zero-temperature $k$-resolved single particle Green function, defined from Eq.~\ref{eq:correlation_function} with ${\hat V}={\hat A}=\hat{c}^{\dagger}_{k\downarrow}$.
The corresponding spectral function, $A_1(k,\omega) = -\frac{1}{\pi} \Im[G(k, \omega)]$, defines the bandstructure of the material. 
The initial walker distribution is given by the perturbed ground state, $\hat{c}^{\dagger}_k | \Psi_0 \ket$, where $| \Psi_0 \ket$ is obtained from a prior ground-state FCIQMC calculation. This starting wavefunction ensures that on average the component of a particular eigenstate in any imaginary-time snapshot is proportional to its transition amplitude in the correlation function.
This approach works particularly well for spectra dominated by a small number of states with large transition amplitudes. Because the transformation to the \Lowdin basis introduces large errors if many states are kept (due to small overlap eigenvalues), we typically limit the number of \Lowdin vectors to between $10$ and $20$, which limits the resolution of the spectrum. Furthermore, high-energy states die away rapidly in the Krylov vectors and so there tends to be significant stochastic errors associated with the calculation of such states. Although this limits the accuracy of KP-FCIQMC over a large energy range, we find that the method is capable of producing accurate spectra in the critical low-energy region and can accurately capture important features such as bandgaps.

\begin{figure}
\includegraphics{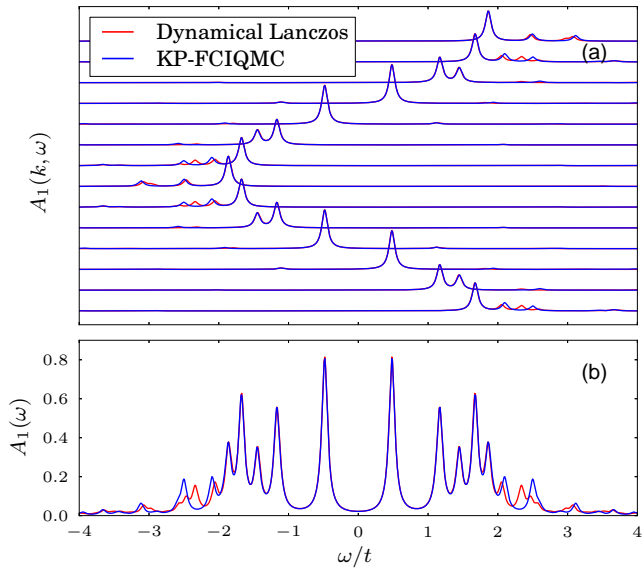}
\caption{(a) $A_1(k,\omega)$ from $k=-\frac{6}{7}\pi$ (bottom) to $k=\pi$ (top) for the 1D 14-site Hubbard model at $U/t=2$, compared to dynamical Lanczos. Poles coming from the ground state or low-lying excited states with large transition amplitudes are captured accurately. (b) The local density of states. The low-energy results are reproduced accurately by KP-FCIQMC while the qualitative behaviour is captured at high energies. Simulation parameters were $\tau=0.01$, $\nadd=3.0$, and a deterministic space of $50,000$ determinants\cite{Petruzielo2012}.}
\label{fig:1_particle_spectra}
\end{figure}

Figure~\ref{fig:1_particle_spectra}(a) presents $A_1(k,\omega)$ for the 14-site Hubbard model at $U/t=2$ with $\sim10^5$ walkers, with $\boldsymbol{S}$ and $\boldsymbol{T}$ averaged over $10$ repeats. 
$35$ Krylov vectors were sampled and $10$ \Lowdin vectors were retained. A complete calculation for a given $k$-sector typically took only $\sim6$ core hours. 
The results are compared to highly-accurate dynamical Lanczos results, using $100$ Lanczos vectors. Figure~\ref{fig:1_particle_spectra}(b) presents the local density of states, computed from the results in (a) via $A(\omega) = \frac{1}{N}\sum_k A(k,\omega)$. The KP-FCIQMC results give high accuracy for low-energy features, with sum rules and causality conditions exactly fulfilled. 
Errors on individual poles can be estimated by repeating results. By comparing eigenvalue estimates from $10$ independent calculations, the bandgap was estimated as $0.96456(14) t$ compared to the exact value of $0.96378 t$. 

\begin{figure}
\includegraphics{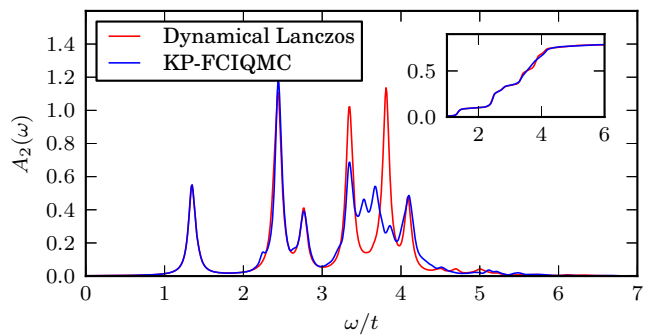}
\caption{$A_2(\omega)$ calculated for the 10-site Hubbard model with $U/t=1$, and compared to near-exact dynamical Lanczos. Inset shows integrated weight, $\int_0^{\omega} A_2(\omega') d\omega'$. Simulation parameters were $\tau=0.01$, with a deterministic space of double excitations.}
\label{fig:2_particle_spectra}
\end{figure}

We also consider the $s$-wave pair-pair dynamic correlation function, a two-body response property of significant relevance in the detection of superconducting quasiparticles. ${\hat V}$ is defined 
by the singlet pairing operator, $\Delta_i=\frac{1}{\sqrt{2}} (c_{i\uparrow} c_{i+1,\downarrow} - c_{i \downarrow} c_{i+1, \uparrow})$, with ${\hat A} = {\hat V}$.
In Fig.~\ref{fig:2_particle_spectra} we present results for this pairing spectrum 
($A_2(\omega)$) for the 10-site Hubbard model at $U/t=1$, 
by computing all $k$-space contributions.
The number of walkers was typically between $10^3$ and $10^4$. 
The initiator adaptation was not applied because the walker population is above the plateau\cite{Spencer2012} height for this system. 
No averaging of $\boldsymbol{T}$ or $\boldsymbol{S}$ over repeated calculations was performed. 
Once again, it is found that low-energy features are calculated accurately, but the quality decreases for higher energy regions of the spectrum.

\emph{Isolated excited states:-} As a further application to larger-scale {\em ab initio} systems, we consider the all-electron ground and first excited state of Neon, in aug-cc-pVDZ and aug-cc-pVTZ basis sets\cite{Dunning}. We work in spaces where $M_s$ is constrained to be zero, but the total spin, $S^2$, is not. The $S=0$ and $S=1$ states are therefore both contained within the same symmetry sector, and the spin-gap can be directly targeted with KP-FCIQMC. The determinant space sizes with these two basis sets are $\sim1.4\times10^8$ and $\sim2.3\times10^{11}$, respectively.
In order to ensure large components of the desired states in the sampled Krylov vectors, the initial wavefunction was created from a linear combination of trial estimates of the ground and first excited states at the inexpensive CISD level of theory.

KP-FCIQMC results are presented in table~\ref{tab:ne_results}, with density matrix renormalisation group (DMRG) results for comparison. DMRG is a highly accurate algorithm, which can also be extended to thermal and spectral quantities, and so is a suitable choice for comparison\cite{DMRG,tDMRG,FTMPS}. KP-FCIQMC results and errors were estimated by averaging over $10$ independent calculations. For the aug-cc-pVDZ results, $2\times10^5$ walkers were used, while $2\times10^6$ walkers were used for the aug-cc-pVTZ basis, with each calculation taking around $100$ core hours for this larger basis. Calculations used 35 Krylov vectors, with 10 \Lowdin vectors retained, providing excellent agreement with DMRG.

\begin{table}
\begin{center}
{\footnotesize
\begin{tabular}{@{\extracolsep{4pt}}ccccc@{}}
\hline
\hline
& \multicolumn{2}{c}{$S=0$} & \multicolumn{2}{c}{$S=1$} \\
\cline{2-3} \cline{4-5}
Basis set & KP-FCIQMC & DMRG & KP-FCIQMC & DMRG \\
\hline
aug-cc-pVDZ & -128.71143(4) & -128.71147 & -127.97787(5) & -127.97794 \\
aug-cc-pVTZ & -128.8258(1) & -128.82514 & -128.109(1) & -128.10919 \\
\hline
\hline
\end{tabular}
}
\caption{Results for the ground ($S=0$) and first excited ($S=1$) states of the Ne atom ($E_h$), 
comparing KP-FCIQMC with DMRG (using M=500 spin-adapted renormalised states for the larger basis)\cite{Block,SpinAdDMRG}. $\tau=0.001$, $n_a=3$ and a deterministic space of single and double excitations.}
\label{tab:ne_results}
\end{center}
\end{table}

\emph{Conclusion:-} We have presented a novel approach to the calculation of excited state, spectral and thermal properties within the FCIQMC framework. In this approach the full Hamiltonian eigenvalue problem is projected into a stochastically sampled Krylov subspace, thus allowing finite-temperature and dynamical quantities to be calculated. 
Since the method exploits sparsity in the sampled wavefunctions, the stochastic dynamic avoids storing Krylov vectors in their entirety, rendering the approach scalable to systems sizes outside the range of the Lanczos method, although in practice this is likely to be restricted if attempting to probe high frequency spectral features.

\section{Acknowledgements}

G.H.B gratefully acknowledges funding from the Royal Society. N.S.B acknowledges Trinity College, Cambridge for funding. This work has been supported by the EPSRC under grant no. EP/J003867/1.


\begin{thebibliography}{39}%
\makeatletter
\providecommand \@ifxundefined [1]{%
 \@ifx{#1\undefined}
}%
\providecommand \@ifnum [1]{%
 \ifnum #1\expandafter \@firstoftwo
 \else \expandafter \@secondoftwo
 \fi
}%
\providecommand \@ifx [1]{%
 \ifx #1\expandafter \@firstoftwo
 \else \expandafter \@secondoftwo
 \fi
}%
\providecommand \natexlab [1]{#1}%
\providecommand \enquote  [1]{``#1''}%
\providecommand \bibnamefont  [1]{#1}%
\providecommand \bibfnamefont [1]{#1}%
\providecommand \citenamefont [1]{#1}%
\providecommand \href@noop [0]{\@secondoftwo}%
\providecommand \href [0]{\begingroup \@sanitize@url \@href}%
\providecommand \@href[1]{\@@startlink{#1}\@@href}%
\providecommand \@@href[1]{\endgroup#1\@@endlink}%
\providecommand \@sanitize@url [0]{\catcode `\\12\catcode `\$12\catcode
  `\&12\catcode `\#12\catcode `\^12\catcode `\_12\catcode `\%12\relax}%
\providecommand \@@startlink[1]{}%
\providecommand \@@endlink[0]{}%
\providecommand \url  [0]{\begingroup\@sanitize@url \@url }%
\providecommand \@url [1]{\endgroup\@href {#1}{\urlprefix }}%
\providecommand \urlprefix  [0]{URL }%
\providecommand \Eprint [0]{\href }%
\providecommand \doibase [0]{http://dx.doi.org/}%
\providecommand \selectlanguage [0]{\@gobble}%
\providecommand \bibinfo  [0]{\@secondoftwo}%
\providecommand \bibfield  [0]{\@secondoftwo}%
\providecommand \translation [1]{[#1]}%
\providecommand \BibitemOpen [0]{}%
\providecommand \bibitemStop [0]{}%
\providecommand \bibitemNoStop [0]{.\EOS\space}%
\providecommand \EOS [0]{\spacefactor3000\relax}%
\providecommand \BibitemShut  [1]{\csname bibitem#1\endcsname}%
\let\auto@bib@innerbib\@empty
\bibitem [{\citenamefont {Foulkes}\ \emph {et~al.}(2001)\citenamefont
  {Foulkes}, \citenamefont {Mitas}, \citenamefont {Needs},\ and\ \citenamefont
  {Rajagopal}}]{Foulkes2001}%
  \BibitemOpen
  \bibfield  {author} {\bibinfo {author} {\bibfnamefont {W.~M.~C.}\
  \bibnamefont {Foulkes}}, \bibinfo {author} {\bibfnamefont {L.}~\bibnamefont
  {Mitas}}, \bibinfo {author} {\bibfnamefont {R.~J.}\ \bibnamefont {Needs}}, \
  and\ \bibinfo {author} {\bibfnamefont {G.}~\bibnamefont {Rajagopal}},\
  }\href@noop {} {\bibfield  {journal} {\bibinfo  {journal} {Rev. Mod. Phys.}\
  }\textbf {\bibinfo {volume} {73}},\ \bibinfo {pages} {33} (\bibinfo {year}
  {2001})}\BibitemShut {NoStop}%
\bibitem [{\citenamefont {Booth}\ \emph {et~al.}(2013)\citenamefont {Booth},
  \citenamefont {Grueneis}, \citenamefont {Kresse},\ and\ \citenamefont
  {Alavi}}]{Booth2013}%
  \BibitemOpen
  \bibfield  {author} {\bibinfo {author} {\bibfnamefont {G.~H.}\ \bibnamefont
  {Booth}}, \bibinfo {author} {\bibfnamefont {A.}~\bibnamefont {Grueneis}},
  \bibinfo {author} {\bibfnamefont {G.}~\bibnamefont {Kresse}}, \ and\ \bibinfo
  {author} {\bibfnamefont {A.}~\bibnamefont {Alavi}},\ }\href@noop {}
  {\bibfield  {journal} {\bibinfo  {journal} {Nature}\ }\textbf {\bibinfo
  {volume} {493}},\ \bibinfo {pages} {365} (\bibinfo {year}
  {2013})}\BibitemShut {NoStop}%
\bibitem [{\citenamefont {Sharma}\ \emph {et~al.}(2014)\citenamefont {Sharma},
  \citenamefont {Yanai}, \citenamefont {Booth}, \citenamefont {Umrigar},\ and\
  \citenamefont {Chan}}]{Sharma2014}%
  \BibitemOpen
  \bibfield  {author} {\bibinfo {author} {\bibfnamefont {S.}~\bibnamefont
  {Sharma}}, \bibinfo {author} {\bibfnamefont {T.}~\bibnamefont {Yanai}},
  \bibinfo {author} {\bibfnamefont {G.~H.}\ \bibnamefont {Booth}}, \bibinfo
  {author} {\bibfnamefont {C.~J.}\ \bibnamefont {Umrigar}}, \ and\ \bibinfo
  {author} {\bibfnamefont {G.~K.-L.}\ \bibnamefont {Chan}},\ }\href@noop {}
  {\bibfield  {journal} {\bibinfo  {journal} {J. Chem. Phys.}\ }\textbf
  {\bibinfo {volume} {140}},\ \bibinfo {pages} {104112} (\bibinfo {year}
  {2014})}\BibitemShut {NoStop}%
\bibitem [{\citenamefont {Virgus}\ \emph {et~al.}(2012)\citenamefont {Virgus},
  \citenamefont {Purwanto}, \citenamefont {Krakauer},\ and\ \citenamefont
  {Zhang}}]{Zhang2012}%
  \BibitemOpen
  \bibfield  {author} {\bibinfo {author} {\bibfnamefont {Y.}~\bibnamefont
  {Virgus}}, \bibinfo {author} {\bibfnamefont {W.}~\bibnamefont {Purwanto}},
  \bibinfo {author} {\bibfnamefont {H.}~\bibnamefont {Krakauer}}, \ and\
  \bibinfo {author} {\bibfnamefont {S.}~\bibnamefont {Zhang}},\ }\href@noop {}
  {\bibfield  {journal} {\bibinfo  {journal} {Phys. Rev. B}\ }\textbf {\bibinfo
  {volume} {86}},\ \bibinfo {pages} {241406} (\bibinfo {year}
  {2012})}\BibitemShut {NoStop}%
\bibitem [{\citenamefont {Sorella}\ \emph {et~al.}(2012)\citenamefont
  {Sorella}, \citenamefont {Otsuka},\ and\ \citenamefont
  {Yunoki}}]{SorellaSpinLiquid}%
  \BibitemOpen
  \bibfield  {author} {\bibinfo {author} {\bibfnamefont {S.}~\bibnamefont
  {Sorella}}, \bibinfo {author} {\bibfnamefont {Y.}~\bibnamefont {Otsuka}}, \
  and\ \bibinfo {author} {\bibfnamefont {S.}~\bibnamefont {Yunoki}},\
  }\href@noop {} {\bibfield  {journal} {\bibinfo  {journal} {Sci. Reps.}\
  }\textbf {\bibinfo {volume} {2}},\ \bibinfo {pages} {992} (\bibinfo {year}
  {2012})}\BibitemShut {NoStop}%
\bibitem [{\citenamefont {White}\ \emph {et~al.}(1989)\citenamefont {White},
  \citenamefont {Scalapino}, \citenamefont {Sugar},\ and\ \citenamefont
  {Bickers}}]{White1989_2}%
  \BibitemOpen
  \bibfield  {author} {\bibinfo {author} {\bibfnamefont {S.~R.}\ \bibnamefont
  {White}}, \bibinfo {author} {\bibfnamefont {D.~J.}\ \bibnamefont
  {Scalapino}}, \bibinfo {author} {\bibfnamefont {R.~L.}\ \bibnamefont
  {Sugar}}, \ and\ \bibinfo {author} {\bibfnamefont {N.~E.}\ \bibnamefont
  {Bickers}},\ }\href@noop {} {\bibfield  {journal} {\bibinfo  {journal} {Phys.
  Rev. Lett.}\ }\textbf {\bibinfo {volume} {63}},\ \bibinfo {pages} {1523}
  (\bibinfo {year} {1989})}\BibitemShut {NoStop}%
\bibitem [{\citenamefont {Motta}\ \emph {et~al.}(2014)\citenamefont {Motta},
  \citenamefont {Galli}, \citenamefont {Moroni},\ and\ \citenamefont
  {Vitali}}]{Motta2014}%
  \BibitemOpen
  \bibfield  {author} {\bibinfo {author} {\bibfnamefont {M.}~\bibnamefont
  {Motta}}, \bibinfo {author} {\bibfnamefont {D.~E.}\ \bibnamefont {Galli}},
  \bibinfo {author} {\bibfnamefont {S.}~\bibnamefont {Moroni}}, \ and\ \bibinfo
  {author} {\bibfnamefont {E.}~\bibnamefont {Vitali}},\ }\href@noop {}
  {\bibfield  {journal} {\bibinfo  {journal} {J. Chem. Phys.}\ }\textbf
  {\bibinfo {volume} {140}},\ \bibinfo {pages} {024107} (\bibinfo {year}
  {2014})}\BibitemShut {NoStop}%
\bibitem [{\citenamefont {Sandvik}\ and\ \citenamefont
  {Kurkijarvi}(1991)}]{SSE}%
  \BibitemOpen
  \bibfield  {author} {\bibinfo {author} {\bibfnamefont {A.~W.}\ \bibnamefont
  {Sandvik}}\ and\ \bibinfo {author} {\bibfnamefont {J.}~\bibnamefont
  {Kurkijarvi}},\ }\href@noop {} {\bibfield  {journal} {\bibinfo  {journal}
  {Phys. Rev. B}\ }\textbf {\bibinfo {volume} {43}},\ \bibinfo {pages} {5950}
  (\bibinfo {year} {1991})}\BibitemShut {NoStop}%
\bibitem [{\citenamefont {Jarrell}\ and\ \citenamefont
  {Gubernatis}(1996)}]{Jarrell1996}%
  \BibitemOpen
  \bibfield  {author} {\bibinfo {author} {\bibfnamefont {M.}~\bibnamefont
  {Jarrell}}\ and\ \bibinfo {author} {\bibfnamefont {J.~E.}\ \bibnamefont
  {Gubernatis}},\ }\href@noop {} {\bibfield  {journal} {\bibinfo  {journal}
  {Phys. Rep.}\ }\textbf {\bibinfo {volume} {269}},\ \bibinfo {pages} {133}
  (\bibinfo {year} {1996})}\BibitemShut {NoStop}%
\bibitem [{\citenamefont {Gull}\ \emph {et~al.}(2007)\citenamefont {Gull},
  \citenamefont {Werner}, \citenamefont {Millis},\ and\ \citenamefont
  {Troyer}}]{Gull07}%
  \BibitemOpen
  \bibfield  {author} {\bibinfo {author} {\bibfnamefont {E.}~\bibnamefont
  {Gull}}, \bibinfo {author} {\bibfnamefont {P.}~\bibnamefont {Werner}},
  \bibinfo {author} {\bibfnamefont {A.}~\bibnamefont {Millis}}, \ and\ \bibinfo
  {author} {\bibfnamefont {M.}~\bibnamefont {Troyer}},\ }\href@noop {}
  {\bibfield  {journal} {\bibinfo  {journal} {Phys. Rev. B}\ }\textbf {\bibinfo
  {volume} {76}},\ \bibinfo {pages} {235123} (\bibinfo {year}
  {2007})}\BibitemShut {NoStop}%
\bibitem [{\citenamefont {Williamson}\ \emph {et~al.}(1998)\citenamefont
  {Williamson}, \citenamefont {Hood}, \citenamefont {Needs},\ and\
  \citenamefont {Rajagopal}}]{Williamson1998}%
  \BibitemOpen
  \bibfield  {author} {\bibinfo {author} {\bibfnamefont {A.~J.}\ \bibnamefont
  {Williamson}}, \bibinfo {author} {\bibfnamefont {R.~Q.}\ \bibnamefont
  {Hood}}, \bibinfo {author} {\bibfnamefont {R.~J.}\ \bibnamefont {Needs}}, \
  and\ \bibinfo {author} {\bibfnamefont {G.}~\bibnamefont {Rajagopal}},\
  }\href@noop {} {\bibfield  {journal} {\bibinfo  {journal} {Phys. Rev. B}\
  }\textbf {\bibinfo {volume} {57}},\ \bibinfo {pages} {12140} (\bibinfo {year}
  {1998})}\BibitemShut {NoStop}%
\bibitem [{\citenamefont {Ceperley}\ and\ \citenamefont
  {Bernu}(1988)}]{Ceperley1988}%
  \BibitemOpen
  \bibfield  {author} {\bibinfo {author} {\bibfnamefont {D.~M.}\ \bibnamefont
  {Ceperley}}\ and\ \bibinfo {author} {\bibfnamefont {B.}~\bibnamefont
  {Bernu}},\ }\href@noop {} {\bibfield  {journal} {\bibinfo  {journal} {J.
  Chem. Phys.}\ }\textbf {\bibinfo {volume} {89}},\ \bibinfo {pages} {6316}
  (\bibinfo {year} {1988})}\BibitemShut {NoStop}%
\bibitem [{\citenamefont {Ten-no}(2013)}]{Tenno2013}%
  \BibitemOpen
  \bibfield  {author} {\bibinfo {author} {\bibfnamefont {S.}~\bibnamefont
  {Ten-no}},\ }\href@noop {} {\bibfield  {journal} {\bibinfo  {journal} {J.
  Chem. Phys.}\ }\textbf {\bibinfo {volume} {138}},\ \bibinfo {pages} {164126}
  (\bibinfo {year} {2013})}\BibitemShut {NoStop}%
\bibitem [{\citenamefont {Booth}\ and\ \citenamefont {Chan}(2012)}]{Chan_H2}%
  \BibitemOpen
  \bibfield  {author} {\bibinfo {author} {\bibfnamefont {G.~H.}\ \bibnamefont
  {Booth}}\ and\ \bibinfo {author} {\bibfnamefont {G.~K.-L.}\ \bibnamefont
  {Chan}},\ }\href@noop {} {\bibfield  {journal} {\bibinfo  {journal} {J. Chem.
  Phys.}\ }\textbf {\bibinfo {volume} {137}},\ \bibinfo {pages} {191102}
  (\bibinfo {year} {2012})}\BibitemShut {NoStop}%
\bibitem [{\citenamefont {Booth}\ \emph {et~al.}(2009)\citenamefont {Booth},
  \citenamefont {Thom},\ and\ \citenamefont {Alavi}}]{Booth2009}%
  \BibitemOpen
  \bibfield  {author} {\bibinfo {author} {\bibfnamefont {G.~H.}\ \bibnamefont
  {Booth}}, \bibinfo {author} {\bibfnamefont {A.~J.~W.}\ \bibnamefont {Thom}},
  \ and\ \bibinfo {author} {\bibfnamefont {A.}~\bibnamefont {Alavi}},\
  }\href@noop {} {\bibfield  {journal} {\bibinfo  {journal} {J. Chem. Phys.}\
  }\textbf {\bibinfo {volume} {131}},\ \bibinfo {pages} {054106} (\bibinfo
  {year} {2009})}\BibitemShut {NoStop}%
\bibitem [{\citenamefont {Spencer}\ \emph {et~al.}(2012)\citenamefont
  {Spencer}, \citenamefont {Blunt},\ and\ \citenamefont
  {Foulkes}}]{Spencer2012}%
  \BibitemOpen
  \bibfield  {author} {\bibinfo {author} {\bibfnamefont {J.~S.}\ \bibnamefont
  {Spencer}}, \bibinfo {author} {\bibfnamefont {N.~S.}\ \bibnamefont {Blunt}},
  \ and\ \bibinfo {author} {\bibfnamefont {W.~M.~C.}\ \bibnamefont {Foulkes}},\
  }\href@noop {} {\bibfield  {journal} {\bibinfo  {journal} {J. Chem. Phys.}\
  }\textbf {\bibinfo {volume} {136}},\ \bibinfo {pages} {054110} (\bibinfo
  {year} {2012})}\BibitemShut {NoStop}%
\bibitem [{\citenamefont {Jaklic}\ and\ \citenamefont
  {Prelovsek}(1994)}]{FTLM}%
  \BibitemOpen
  \bibfield  {author} {\bibinfo {author} {\bibfnamefont {J.}~\bibnamefont
  {Jaklic}}\ and\ \bibinfo {author} {\bibfnamefont {P.}~\bibnamefont
  {Prelovsek}},\ }\href@noop {} {\bibfield  {journal} {\bibinfo  {journal}
  {Phys. Rev. B}\ }\textbf {\bibinfo {volume} {49}},\ \bibinfo {pages} {5065}
  (\bibinfo {year} {1994})}\BibitemShut {NoStop}%
\bibitem [{\citenamefont {Prelov\v{s}ek}\ and\ \citenamefont
  {Bon\v{c}a}(2013)}]{Prelovsek2013}%
  \BibitemOpen
  \bibfield  {author} {\bibinfo {author} {\bibfnamefont {P.}~\bibnamefont
  {Prelov\v{s}ek}}\ and\ \bibinfo {author} {\bibfnamefont {J.}~\bibnamefont
  {Bon\v{c}a}},\ }\href@noop {} {\emph {\bibinfo {title} {Strongly Correlated
  Systems}}}\ (\bibinfo  {publisher} {Springer Berlin Heidelberg},\ \bibinfo
  {year} {2013})\ pp.\ \bibinfo {pages} {1--30}\BibitemShut {NoStop}%
\bibitem [{\citenamefont {Dargel}\ \emph {et~al.}(2012)\citenamefont {Dargel},
  \citenamefont {Woellert}, \citenamefont {Honecker}, \citenamefont
  {McCulloch}, \citenamefont {Schollwoeck},\ and\ \citenamefont
  {Pruschke}}]{MPSLanc}%
  \BibitemOpen
  \bibfield  {author} {\bibinfo {author} {\bibfnamefont {P.~E.}\ \bibnamefont
  {Dargel}}, \bibinfo {author} {\bibfnamefont {A.}~\bibnamefont {Woellert}},
  \bibinfo {author} {\bibfnamefont {A.}~\bibnamefont {Honecker}}, \bibinfo
  {author} {\bibfnamefont {I.~P.}\ \bibnamefont {McCulloch}}, \bibinfo {author}
  {\bibfnamefont {U.}~\bibnamefont {Schollwoeck}}, \ and\ \bibinfo {author}
  {\bibfnamefont {T.}~\bibnamefont {Pruschke}},\ }\href@noop {} {\bibfield
  {journal} {\bibinfo  {journal} {Phys. Rev. B}\ }\textbf {\bibinfo {volume}
  {85}},\ \bibinfo {pages} {205119} (\bibinfo {year} {2012})}\BibitemShut
  {NoStop}%
\bibitem [{\citenamefont {Hams}\ and\ \citenamefont
  {De~Raedt}(2000)}]{Hams2000}%
  \BibitemOpen
  \bibfield  {author} {\bibinfo {author} {\bibfnamefont {A.}~\bibnamefont
  {Hams}}\ and\ \bibinfo {author} {\bibfnamefont {H.}~\bibnamefont
  {De~Raedt}},\ }\href@noop {} {\bibfield  {journal} {\bibinfo  {journal}
  {Phys. Rev. E}\ }\textbf {\bibinfo {volume} {62}},\ \bibinfo {pages} {4365}
  (\bibinfo {year} {2000})}\BibitemShut {NoStop}%
\bibitem [{\citenamefont {Sugiura}\ and\ \citenamefont
  {Shimizu}(2013)}]{Shimizu2013}%
  \BibitemOpen
  \bibfield  {author} {\bibinfo {author} {\bibfnamefont {S.}~\bibnamefont
  {Sugiura}}\ and\ \bibinfo {author} {\bibfnamefont {A.}~\bibnamefont
  {Shimizu}},\ }\href@noop {} {\bibfield  {journal} {\bibinfo  {journal} {Phys.
  Rev. Lett.}\ }\textbf {\bibinfo {volume} {111}},\ \bibinfo {pages} {010401}
  (\bibinfo {year} {2013})}\BibitemShut {NoStop}%
\bibitem [{\citenamefont {Shepherd}\ \emph {et~al.}(2014)\citenamefont
  {Shepherd}, \citenamefont {Scuseria},\ and\ \citenamefont
  {Spencer}}]{shepherd2014}%
  \BibitemOpen
  \bibfield  {author} {\bibinfo {author} {\bibfnamefont {J.~J.}\ \bibnamefont
  {Shepherd}}, \bibinfo {author} {\bibfnamefont {G.~E.}\ \bibnamefont
  {Scuseria}}, \ and\ \bibinfo {author} {\bibfnamefont {J.~S.}\ \bibnamefont
  {Spencer}},\ }\href@noop {} {\bibfield  {journal} {\bibinfo  {journal} {Phys.
  Rev. B}\ }\textbf {\bibinfo {volume} {90}},\ \bibinfo {pages} {155130}
  (\bibinfo {year} {2014})}\BibitemShut {NoStop}%
\bibitem [{fn1()}]{fn1}%
  \BibitemOpen
  \href@noop {} {}\bibinfo {note} {In FCIQMC the propagator $\hat{P} =
  \mathbb{1} - \tau (\hat{H} - S\mathbb{1})$ is used, where $\tau$ is a small
  timestep and $S$ is allowed to vary based on the walker population, $\sum_i
  |q_i|$, where $q_i$ is the walker weight on site $i$. See ref.[15] for more
  details.}\BibitemShut {Stop}%
\bibitem [{\citenamefont {Caffarel}\ \emph {et~al.}(1991)\citenamefont
  {Caffarel}, \citenamefont {Gadea},\ and\ \citenamefont
  {Ceperley}}]{Caffarel1991}%
  \BibitemOpen
  \bibfield  {author} {\bibinfo {author} {\bibfnamefont {M.}~\bibnamefont
  {Caffarel}}, \bibinfo {author} {\bibfnamefont {F.~X.}\ \bibnamefont {Gadea}},
  \ and\ \bibinfo {author} {\bibfnamefont {D.~M.}\ \bibnamefont {Ceperley}},\
  }\href@noop {} {\bibfield  {journal} {\bibinfo  {journal} {Europhys. Lett.}\
  }\textbf {\bibinfo {volume} {16}},\ \bibinfo {pages} {249} (\bibinfo {year}
  {1991})}\BibitemShut {NoStop}%
\bibitem [{\citenamefont {Zhang}\ and\ \citenamefont
  {Kalos}(1993)}]{Zhang1993}%
  \BibitemOpen
  \bibfield  {author} {\bibinfo {author} {\bibfnamefont {S.}~\bibnamefont
  {Zhang}}\ and\ \bibinfo {author} {\bibfnamefont {M.~H.}\ \bibnamefont
  {Kalos}},\ }\href@noop {} {\bibfield  {journal} {\bibinfo  {journal} {J.
  Stat. Phys.}\ }\textbf {\bibinfo {volume} {70}},\ \bibinfo {pages} {515}
  (\bibinfo {year} {1993})}\BibitemShut {NoStop}%
\bibitem [{\citenamefont {Blunt}\ \emph {et~al.}(2014)\citenamefont {Blunt},
  \citenamefont {Rogers}, \citenamefont {Spencer},\ and\ \citenamefont
  {Foulkes}}]{Blunt2014}%
  \BibitemOpen
  \bibfield  {author} {\bibinfo {author} {\bibfnamefont {N.~S.}\ \bibnamefont
  {Blunt}}, \bibinfo {author} {\bibfnamefont {T.~W.}\ \bibnamefont {Rogers}},
  \bibinfo {author} {\bibfnamefont {J.~S.}\ \bibnamefont {Spencer}}, \ and\
  \bibinfo {author} {\bibfnamefont {W.~M.~C.}\ \bibnamefont {Foulkes}},\
  }\href@noop {} {\bibfield  {journal} {\bibinfo  {journal} {Phys. Rev. B}\
  }\textbf {\bibinfo {volume} {89}},\ \bibinfo {pages} {245124} (\bibinfo
  {year} {2014})}\BibitemShut {NoStop}%
\bibitem [{\citenamefont {Overy}\ \emph {et~al.}(2014)\citenamefont {Overy},
  \citenamefont {Booth}, \citenamefont {Blunt}, \citenamefont {Shepherd},
  \citenamefont {Cleland},\ and\ \citenamefont {Alavi}}]{Overy2014}%
  \BibitemOpen
  \bibfield  {author} {\bibinfo {author} {\bibfnamefont {C.}~\bibnamefont
  {Overy}}, \bibinfo {author} {\bibfnamefont {G.~H.}\ \bibnamefont {Booth}},
  \bibinfo {author} {\bibfnamefont {N.~S.}\ \bibnamefont {Blunt}}, \bibinfo
  {author} {\bibfnamefont {J.~J.}\ \bibnamefont {Shepherd}}, \bibinfo {author}
  {\bibfnamefont {D.}~\bibnamefont {Cleland}}, \ and\ \bibinfo {author}
  {\bibfnamefont {A.}~\bibnamefont {Alavi}},\ }\href@noop {} {\bibfield
  {journal} {\bibinfo  {journal} {J. Chem. Phys.}\ }\textbf {\bibinfo {volume}
  {141}},\ \bibinfo {pages} {244117} (\bibinfo {year} {2014})}\BibitemShut
  {NoStop}%
\bibitem [{fn2()}]{fn2}%
  \BibitemOpen
  \href@noop {} {}\bibinfo {note} {In practice, we sample the wavefunction with
  a linearly increasing number of iterations between selections, such that
  vectors during early iterations are sampled with finer resolution, where the
  wavefunction is rapidly changing and contains significant projections onto
  excited eigenstates.}\BibitemShut {Stop}%
\bibitem [{\citenamefont {Iitaka}\ and\ \citenamefont
  {Ebisuzaki}(2004)}]{Iitaka2004}%
  \BibitemOpen
  \bibfield  {author} {\bibinfo {author} {\bibfnamefont {T.}~\bibnamefont
  {Iitaka}}\ and\ \bibinfo {author} {\bibfnamefont {T.}~\bibnamefont
  {Ebisuzaki}},\ }\href@noop {} {\bibfield  {journal} {\bibinfo  {journal}
  {Phys. Rev. E}\ }\textbf {\bibinfo {volume} {69}},\ \bibinfo {pages} {057701}
  (\bibinfo {year} {2004})}\BibitemShut {NoStop}%
\bibitem [{\citenamefont {Schnack}\ and\ \citenamefont
  {Wendland}(2010)}]{Schnack2010}%
  \BibitemOpen
  \bibfield  {author} {\bibinfo {author} {\bibfnamefont {J.}~\bibnamefont
  {Schnack}}\ and\ \bibinfo {author} {\bibfnamefont {O.}~\bibnamefont
  {Wendland}},\ }\href@noop {} {\bibfield  {journal} {\bibinfo  {journal} {Eur.
  Phys. J. B.}\ }\textbf {\bibinfo {volume} {78}},\ \bibinfo {pages} {535}
  (\bibinfo {year} {2010})}\BibitemShut {NoStop}%
\bibitem [{\citenamefont {Hanebaum}\ and\ \citenamefont
  {Schnack}(2014)}]{Hanebaum2014}%
  \BibitemOpen
  \bibfield  {author} {\bibinfo {author} {\bibfnamefont {O.}~\bibnamefont
  {Hanebaum}}\ and\ \bibinfo {author} {\bibfnamefont {J.}~\bibnamefont
  {Schnack}},\ }\href@noop {} {\bibfield  {journal} {\bibinfo  {journal} {Eur.
  Phys. J. B.}\ }\textbf {\bibinfo {volume} {3}},\ \bibinfo {pages} {194}
  (\bibinfo {year} {2014})}\BibitemShut {NoStop}%
\bibitem [{\citenamefont {Petruzielo}\ \emph {et~al.}(2012)\citenamefont
  {Petruzielo}, \citenamefont {Holmes}, \citenamefont {Changlani},
  \citenamefont {Nightingale},\ and\ \citenamefont {Umrigar}}]{Petruzielo2012}%
  \BibitemOpen
  \bibfield  {author} {\bibinfo {author} {\bibfnamefont {F.~R.}\ \bibnamefont
  {Petruzielo}}, \bibinfo {author} {\bibfnamefont {A.~A.}\ \bibnamefont
  {Holmes}}, \bibinfo {author} {\bibfnamefont {H.~J.}\ \bibnamefont
  {Changlani}}, \bibinfo {author} {\bibfnamefont {M.~P.}\ \bibnamefont
  {Nightingale}}, \ and\ \bibinfo {author} {\bibfnamefont {C.~J.}\ \bibnamefont
  {Umrigar}},\ }\href@noop {} {\bibfield  {journal} {\bibinfo  {journal} {Phys.
  Rev. Lett.}\ }\textbf {\bibinfo {volume} {109}},\ \bibinfo {pages} {230201}
  (\bibinfo {year} {2012})}\BibitemShut {NoStop}%
\bibitem [{\citenamefont {Cleland}\ \emph {et~al.}(2011)\citenamefont
  {Cleland}, \citenamefont {Booth},\ and\ \citenamefont {Alavi}}]{EAs_iFCIQMC}%
  \BibitemOpen
  \bibfield  {author} {\bibinfo {author} {\bibfnamefont {D.~M.}\ \bibnamefont
  {Cleland}}, \bibinfo {author} {\bibfnamefont {G.~H.}\ \bibnamefont {Booth}},
  \ and\ \bibinfo {author} {\bibfnamefont {A.}~\bibnamefont {Alavi}},\
  }\href@noop {} {\bibfield  {journal} {\bibinfo  {journal} {J. Chem. Phys.}\
  }\textbf {\bibinfo {volume} {134}},\ \bibinfo {pages} {024112} (\bibinfo
  {year} {2011})}\BibitemShut {NoStop}%
\bibitem [{\citenamefont {Dunning}(1989)}]{Dunning}%
  \BibitemOpen
  \bibfield  {author} {\bibinfo {author} {\bibfnamefont {T.}~\bibnamefont
  {Dunning}},\ }\href@noop {} {\bibfield  {journal} {\bibinfo  {journal} {J.
  Chem. Phys.}\ }\textbf {\bibinfo {volume} {90}},\ \bibinfo {pages} {1007}
  (\bibinfo {year} {1989})}\BibitemShut {NoStop}%
\bibitem [{\citenamefont {White}(1992)}]{DMRG}%
  \BibitemOpen
  \bibfield  {author} {\bibinfo {author} {\bibfnamefont {S.~R.}\ \bibnamefont
  {White}},\ }\href@noop {} {\bibfield  {journal} {\bibinfo  {journal} {Phys.
  Rev. Lett.}\ }\textbf {\bibinfo {volume} {69}},\ \bibinfo {pages} {2863}
  (\bibinfo {year} {1992})}\BibitemShut {NoStop}%
\bibitem [{\citenamefont {White}\ and\ \citenamefont {Feiguin}(2004)}]{tDMRG}%
  \BibitemOpen
  \bibfield  {author} {\bibinfo {author} {\bibfnamefont {S.~R.}\ \bibnamefont
  {White}}\ and\ \bibinfo {author} {\bibfnamefont {A.~E.}\ \bibnamefont
  {Feiguin}},\ }\href@noop {} {\bibfield  {journal} {\bibinfo  {journal} {Phys.
  Rev. Lett.}\ }\textbf {\bibinfo {volume} {93}},\ \bibinfo {pages} {076401}
  (\bibinfo {year} {2004})}\BibitemShut {NoStop}%
\bibitem [{\citenamefont {Feiguin}\ and\ \citenamefont {White}(2005)}]{FTMPS}%
  \BibitemOpen
  \bibfield  {author} {\bibinfo {author} {\bibfnamefont {A.~E.}\ \bibnamefont
  {Feiguin}}\ and\ \bibinfo {author} {\bibfnamefont {S.~R.}\ \bibnamefont
  {White}},\ }\href@noop {} {\bibfield  {journal} {\bibinfo  {journal} {Phys.
  Rev. B}\ }\textbf {\bibinfo {volume} {72}},\ \bibinfo {pages} {220401(R)}
  (\bibinfo {year} {2005})}\BibitemShut {NoStop}%
\bibitem [{\citenamefont {Olivares-Amaya}\ \emph {et~al.}(2015)\citenamefont
  {Olivares-Amaya}, \citenamefont {Hu}, \citenamefont {Nakatani}, \citenamefont
  {Sharma}, \citenamefont {Yang},\ and\ \citenamefont {Chan}}]{Block}%
  \BibitemOpen
  \bibfield  {author} {\bibinfo {author} {\bibfnamefont {R.}~\bibnamefont
  {Olivares-Amaya}}, \bibinfo {author} {\bibfnamefont {W.}~\bibnamefont {Hu}},
  \bibinfo {author} {\bibfnamefont {N.}~\bibnamefont {Nakatani}}, \bibinfo
  {author} {\bibfnamefont {S.}~\bibnamefont {Sharma}}, \bibinfo {author}
  {\bibfnamefont {J.}~\bibnamefont {Yang}}, \ and\ \bibinfo {author}
  {\bibfnamefont {G.~K.-L.}\ \bibnamefont {Chan}},\ }\href@noop {} {\bibfield
  {journal} {\bibinfo  {journal} {J. Chem. Phys.}\ }\textbf {\bibinfo {volume}
  {142}},\ \bibinfo {pages} {034102} (\bibinfo {year} {2015})}\BibitemShut
  {NoStop}%
\bibitem [{\citenamefont {Sharma}\ and\ \citenamefont
  {Chan}(2012)}]{SpinAdDMRG}%
  \BibitemOpen
  \bibfield  {author} {\bibinfo {author} {\bibfnamefont {S.}~\bibnamefont
  {Sharma}}\ and\ \bibinfo {author} {\bibfnamefont {G.~K.-L.}\ \bibnamefont
  {Chan}},\ }\href@noop {} {\bibfield  {journal} {\bibinfo  {journal} {J. Chem.
  Phys.}\ }\textbf {\bibinfo {volume} {136}},\ \bibinfo {pages} {124121}
  (\bibinfo {year} {2012})}\BibitemShut {NoStop}%
\end{thebibliography}
%

\end{document}